\documentclass[12pt]{article}

\usepackage{amsfonts,epsfig}

\def\rme{\mbox{e}}
\def\rmd{\mbox{d}}

\begin{document}

\title{Mother wavelet functions generalized through $q$-exponentials}

\author{Ernesto P. Borges$^{a,b}$%
\thanks{E-mail: ernesto@ufba.br, tsallis@cbpf.br, vivas@ufba.br, 
        randrade@ufba.br} ,
Constantino Tsallis$^{b,c*}$,\\
Jos\'e G.~V. Miranda$^{d*}$,
Roberto F.~S. Andrade$^{d*}$
\\ $ $ \\
{\normalsize $^a$Escola Polit\'ecnica, Universidade Federal da Bahia,} \\
{\normalsize R. Aristides Novis, 2, 40210-630 Salvador, BA, Brazil}
\\
{\normalsize $^b$Centro Brasileiro de Pesquisas F\'\i sicas,} \\
{\normalsize R. Dr.  Xavier Sigaud 150, 22290-180 Rio de Janeiro, RJ, Brazil}
\\
{\normalsize $^c$Santa Fe Institute,} 
{\normalsize 1399 Hyde Park Road, Santa Fe, NM, 87501, USA}
\\
{\normalsize $^d$Instituto de F\'\i sica, Universidade Federal da Bahia,} \\
{\normalsize Campus Universit\'ario de Ondina, 40210-340 Salvador, BA, Brazil}
}

\date{}

\maketitle

\begin{abstract}
We generalize some widely used mother wavelets by means of the
$q$-exponen\-tial function $\rme_q^x \equiv [1+(1-q)x]^{1/(1-q)}$
($q \in {\mathbb R}$, $\rme_1^x=\rme^x$) that emerges from
nonextensive statistical mechanics.
Particularly, we define extended versions of the mexican hat and the Morlet
wavelets. We also introduce new wavelets that are $q$-generalizations of the
trigonometric functions.
All cases reduce to the usual ones as $q \rightarrow 1$.
Within nonextensive statistical mechanics, departures from unity of the
entropic index $q$ are expected in the presence of long-range interactions,
long-term memory, multi-fractal structures, among others.
Consistently the analysis of signals associated with such
features is hopefully improved by proper tuning of the value of $q$.
We exemplify with the WTMM Method for mono- and multi-fractal self-affine
signals.
\end{abstract}

\begin{quote}
PACS: 
02.30.-f, 
02.30.Nw, 
05.20.-y, 
05.45.Tp, 
\end{quote}

\section{Introduction}

The basic idea behind the analysis of a temporal or spatial signal by a
Fourier transform is {\it similarity}. For instance, the inner product
$\langle \phi_n(x),f(x) \rangle$ expresses how similar are the function
$f(x)$ and the $n^{\mbox{{\tiny th}}}$ element of the basis $\{\phi_n(x)\}$
(the kernel of the transform). The kernel of Fourier transform is a plane
wave and here lies its simplicity, but also its limitation. Rigorously
speaking, a plane wave exists
everywhere in the universe and at all (past and future) times.
Hence there are no plane waves in nature.
All physical signals are temporary and spatially limited.
As a consequence of the infinitely extended kernel, the Fourier transform
is unable to satisfactorily determine {\it when} a burst has occured,
or {\it where} edges of images are located, for example
(the Gibbs phenomenon, see, for instance, \cite{kreyszig,wylie}).

The first attempt to overcome limitations of Fourier analysis
was that of Gabor \cite{Gabor}, who introduced a windowed Fourier transform
\cite{Kaiser}.
He used a modulation function $g(u-t)$ in order to localize
the signal in a time interval $[t-T,t]$ ($\mbox{supp } g \subset [-T,0]$).
Translations in time would cover the whole signal. With this, he achieved
time-frequency localization. But a problem still remains with the use of a
fixed scale window: signal details much smaller than the window width $T$
are detected, but not localized. They appear in the frequency behaviour of
the windowed Fourier transform in a similar way it would appear in the
usual Fourier transform.
Signal features much larger than $T$, on the other hand, appear in the time
behaviour of the windowed Fourier transform (i.e., they are not detected).

To avoid this problem, it is sufficient a scale free transform.
Wavelet analysis was developed to accomplish this goal.
The windowed Fourier transform uses a fixed size window and fills it with
oscillations of different frequencies (consequently, varies the number of
oscillations within the window). The wavelet transform uses a function
with fixed number of oscillations and vary the width of the window.
Dilations and translations generate a complete analysis of the signal.
This procedure automatically uses small windows to identify high-frequency
components of a signal, and large windows for low-frequency components.
Wavelet analysis provides a time-scale (instead of a time-frequency)
localization.

Wavelet analysis was developed more than a century and a half after
the pioneering work of Fourier \cite{fourier}.
Since the 1980's it has grown so fast that it has already consolidated as a
field of research of its own.

\bigskip

Another emerging science is nonextensive statistical mechanics,
that also began in the 1980's, with the
formulation of the concept of a nonextensive entropy
\cite{CT1988}, that generalizes the Boltzmann-Gibbs entropy, as a basis of a
generalization of standard statistical mechanics itself.
Here, it took also more than a century since the first formulation of the
concept of entropy.
Nonextensive statistical mechanics \cite{CT1988,ct-evaldo,abe-springer,%
ct-gell-mann}
seems to give an underlying formal basis for a variety of complex phenomena,
such as anomalous diffusion
\cite{Alemany:94,Zanette:95,Tsallis:LevyFlights,Zanette:BJP,Wilk:2000,%
CTLevy,CTPrato},
self-gravitating systems \cite{Plastino_Plastino,Hamity},
turbulence \cite{Boghosian:96,Anteneodo,Boghosian:99,Arimitsu_a,Arimitsu_b,%
Beck:2000,Beck_Swinney:2001}
cosmic rays \cite{cosmic-1,cosmic-2},
among others (for un updated bibliography, see \cite{http}).

The definition of the nonextensive $q$-entropy \cite{CT1988}
is
\begin{eqnarray}
\label{Sq}
S_q \equiv k \frac{1-\sum_{i=1}^W p_i^q}{q-1} \qquad (k>0).
\end{eqnarray}
The entropic index $q$ characterizes the generalization and the
Boltzmann-Gibbs-Shanon entropy, $S_1 = -k\sum_i p_i \ln p_i$,
is recovered at $q \rightarrow 1$.
The canonical ensemble is obtained by maximizing (\ref{Sq}) with the
norm constraint $(\sum_{i=1}^W p_i=1)$ and also imposing the constancy of
the normalized $q$-expectation value of the energy \cite{TMP}
\begin{eqnarray}
\label{Uq}
U_q=\sum_{i=1}^W \epsilon_i P_i^{(q)},
\end{eqnarray}
where $\{\epsilon_i\}_{i=1}^W$ are the eigenvalues of the Hamiltonian
of the system, and $\{P_i^{(q)}\}_{i=1}^W$ are the escort probabilities,
defined by \cite{beck}
\begin{eqnarray}
 \label{escort}
 P_i^{(q)} \equiv \frac{p_i^q}{\sum_{j=1}^W p_j^q},
\end{eqnarray}
with $p_i$ being the probability associated with the $i^{\mbox{{\tiny th}}}$ 
state, given by
\begin{eqnarray}
\label{pi}
p_i=\frac{\left[ 1-(1-q) \beta' \epsilon_i \right]^{\frac{1}{1-q}}}{Z_q'}.
\end{eqnarray}
If $q<1$, $p_i \equiv 0$ whenever $[1-(1-q)\beta' \epsilon_i] \le 0$
(cut-off condition).
The partition function $Z'_q$ is defined as
\begin{eqnarray}
\label{Zq}
Z_q' \equiv \sum_{j=1}^W [1-(1-q) \beta' \epsilon_j]^{\frac{1}{1-q}}
\end{eqnarray}
and
\begin{eqnarray}
\label{beta'}
\beta' = \frac{\beta}{\sum_{j=1}^W p_j^q + (1-q) \beta U_q},
\end{eqnarray}
where $\beta$ is the Lagrange parameter associated with the constraint
(\ref{Uq}).
The striking features of (\ref{pi}) are its power-law for $q>1$,
and the cut-off for $q<1$, which generates quite different behaviours when
compared with the Boltzmann-Gibbs exponential distribution.
The meaning of the word {\it nonextensive} can be easily understood if we
consider a system composed of two independent subsystems (in the sense
of probability theory, $p_{ij}^{(A+B)}=p_i^{(A)}p_j^{(B)}$). According
to (\ref{Sq}), the $q$-entropy of the composite system is given by
\begin{eqnarray}
\label{q-additivity}
S_q^{(A+B)}=S_q^{(A)}+S_q^{(B)}+(1-q)S_q^{(A)}S_q^{(B)}.
\end{eqnarray}
Clearly $q \ne 1$ yields nonadditivity, or nonextensivity. A system with
$q>1$ is said to be subadditive, and if $q<1$, it is superadditive.

The functional forms of (\ref{Sq}) and (\ref{pi}) inspired the definition
of generalizations of the logarithm and exponential functions
\cite{CTQuimicaNova,EPB1998}:
\begin{eqnarray}
\label{lnq}
\ln _q x \equiv \frac{x^{1-q}-1}{1-q},
\end{eqnarray}
\begin{eqnarray}
\label{expq}
\rme_q^x \equiv
\left\{
\begin{array}{ll}
\left[1+(1-q)x\right]^\frac{1}{1-q}, & \mbox{if } [1+(1-q)x] > 0 \\
\noalign{\smallskip}
0\;,   & \mbox{otherwise}.
\end{array}
\right.
\end{eqnarray}
It follows immediately that $\ln_q(\rme_q^x) = \rme_q^{\ln_q(x)} =x$.
The original logarithm and exponential functions are recovered
as the particular cases $\ln_1 x$ and $\rme_1^x$.
Many formal developments have been made concerning such functions.
Among them we point out the generalization \cite{qshannon} of the
celebrated Shannon's theorem.
Obviously there are infinitely many ways of generalizing a given
function. One commonly form of generalizing the exponential is
$
\exp_q (x)\equiv \sum_n x^n/(n)_{q}!,
$
with $(n)_{q}!=\prod_{j=1}^{n} (j)_{q}$
and $(j)_{q}=(q^{j}-1)/(q-1)$ and also $(0)_{q}!=1$.
This particular expression have applications in quantum groups
\cite{kassel}.
In this paper we won't explore this possibility, but rather that one
given by
Equation (\ref{expq}) above, that naturally emerges from
nonextensive statistical mechanics.
Relations between wavelet analysis and nonextensive statistical mechanics
have already been reported, including applications in biophysics, as
the analysis of EEG signals
\cite{gamero-plastino:wavelets,capurro-plastino:eeg-1,capurro-plastino:eeg-2,%
martin-plastino:wavelets,tong-eeg}.

The purpose of this paper is to extend the use of such $q$-deformed functions
into wavelet analysis by generalizing some widely used wavelet functions.
Particularly, we generalize the mexican hat (Section~\ref{sec:q-mexicanhat})
and the modulated Gaussian (Section~\ref{sec:q-gaussian}).
Moreover, we introduce in Section~\ref{sec:q-trigonometric} a pair of
even and odd wavelets based on the generalization of trigonometric functions.
Finally (Section~\ref{sec:example}) we exemplify, with the Wavelet Transform
Modulus-Maxima Method for mono- and multi-fractal self-affine signals,
the possible use of the introduced functions,
and, lastly, we present our final remarks (Section~\ref{sec:final}).

\section{$q$-Mexican Hat}
\label{sec:q-mexicanhat}

Before we discuss the properties of the wavelets based on
definitions (\ref{lnq}) and (\ref{expq}), let us indicate some basic
facts about continuous $g$-wavelet transform of a function $f$,
$T_g(a,b)f$, which is defined by
\begin{equation}
\label{eq0} T_g(a,b)f=a^{-1/2}C_g^{-1/2}\int_{-\infty}^{\infty}\rmd x 
\,f(x)\,g(\frac{x-b}a).
\end{equation}
In order that a given function $g\in L^2(\mathbb{R})$ can be considered as 
a mother wavelet, it must satisfy the so-called admissibility
condition, that can be expressed in terms of its Fourier transform
$\hat{g}(\omega)$ by
\begin{equation}
\label{eq11a}C_g=2\pi\int_{-\infty}^{\infty}\rmd\omega
\,|\hat{g}(\omega)|^2/|\omega|<\infty .
\end{equation}
It essentially means that $g(x)$ has zero mean, i.e.,
\begin{equation}
\label{eq11b}\int_{-\infty}^{\infty} \rmd x\,g(x)=0
\end{equation}
and ensures that the the original function $f(x)$ can be recovered
from its wavelet transform as
\begin{equation}
\label{eq11c} f(x)=C_g^{-1/2} \int_{-\infty}^{\infty} \rmd b 
\int_0^{\infty} \rmd a\,\frac{1}{a^2}\,T_g(a,b)f\,g(\frac{x-b}a).
\end{equation}
This expression shows that, if a mother wavelet $g$ satisfies
(\ref{eq11a}), it generates, by its translations and dilations, a
basis in the Hilbert space, so that any function $f\in L^2(\mathbb{R})$ can 
be expressed in terms of its $g-$wavelet components. It is well
known that orthonormal wavelet bases can only be defined within
the discrete formulation, i.e., when $b\in \mathbb{Z}$ and $a=a_0^j$, with
$a_0>0$ and $j\in \mathbb{Z}$. So, orthonormal bases can not be defined
with the help of the functions we discuss below.

A simple and quite common example of continuous wavelet is the mexican hat 
(see, for instance,
\cite{Daubechies90,Daubechies_Ten_Lectures,Hernandez_Weiss}),
that is generated from the Gaussian distribution:
\begin{eqnarray}
\label{sombrero}
\psi(x)&=&-A\,\frac{\rmd^2\,\rme^{-x^2/2}}{\rmd x^2},
\nonumber\\ \noalign{\smallskip}
&=&A\,(1-x^2)\,\rme^{-x^2/2}.
\end{eqnarray}
The normalization constant is given by $A=2/(\pi^{1/4}\sqrt{3})$.
Generalization of Gaussian distribution within nonextensive
scenario we are focusing here has already been made \cite{CTLevy,CTPrato}.
The $q$-Gaussian ($\propto \rme_q^{-\beta x^2}$)
unifies a great variety of different distributions into a
single family, parameterized by $q$ (see \cite{CTLevy,CTPrato} for details):
Gaussian distribution is recovered, of course, for $q=1$.
Moreover, for $q=2$, we have the Cauchy-Lorentz distribution.
Besides, $q \to 3$ yields a completely flat distribution,
and $q \to -\infty$ yields Dirac's $\delta$.
This unifying character of the $q$-Gaussian, as well as its
empirically observed occurence in a variety of complex phenomena
(as cited in the Introdutcion) stimulate us to explore its potential use in
wavelet analysis. One of the more simple applications that immediately come
to our minds is to use it in order to generalize the mexican hat,
a function that is proportional to the second derivative of a Gaussian.
We could simply replace an ordinary Gaussian by a $q$-Gaussian, which is a
legitimate choice. Instead of doing that, we prefer to adopt a variation
of it, and take the second derivative of a {\em power} of a $q$-Gaussian,
according to the recipe
\begin{eqnarray}
\label{qsombrero_def}
\psi_q(x) \propto \frac{\rmd^2\,[\rme_q^{-\beta x^2}]^{2-q}}{\rmd x^2},
\end{eqnarray}
which yields the expression for the $q$-mexican hat
\begin{eqnarray}
\label{qsombrero}
\psi_q(x)=
A_q\,\left[1-(3-q)\beta x^2 \right]\,\left[\rme_q^{-\beta x^2}\right]^q.
\end{eqnarray}
With this choice we arrive, according to the expression
above,
at a $q$-Gaussian raised to the power $q$, resembling the escort probabilities,
Equation (\ref{escort}), widely used in nonextensive formalism \cite{TMP}.
We call the attention of the reader that
$[\rme_q^a]^b\ne[\rme_q^{ab}]$, except, of course, in the particular
case $q=1$.
Because of this, the semigroup property is not satisfied, but it is
exactly this that
generates its nonextensive behaviour Equation (\ref{q-additivity}). 
For some properties of the $q$-exponential and the $q$-logarithm,
see \cite{ct-abe-springer,epb:q-algebra,suyari:errors,%
suyari:mathematicalstructure,suyari:stirling}.

If $-1 < q < 1$ and $|x| = [(1-q)\beta]^{-1/2}$, the cut-off of the
$q$-exponential (see Equation (\ref{expq})) imposes $\psi_q(x) = 0$. It is
easily proven that, for $ -1< q < 3$, Equation (\ref{eq11a}) holds, so
that any $\psi_q(x)$ is a well defined mother wavelet. For $q \leq
-1$ or $ q\geq 3$, $\psi_q(x)$ does not satisfy the admissibility
condition and can not be used to define a useful mother wavelet.

The normalization constant is given by
\begin{eqnarray}
\label{Aq:q>1}
A_q =
\frac{\beta^{1/4}}{\pi^{1/4}\sqrt{3}} \,
\left[\frac{(q-1)^{5/2}\,\Gamma\left(\frac{2q}{q-1}\right)}
{\Gamma\left(\frac{2q}{q-1}-\frac{5}{2}\right)}
\right]^{1/2}
\end{eqnarray}
if $1<q<3$, and
\begin{eqnarray}
\label{Aq:q<1}
A_q =
\frac{\beta^{1/4}}{\pi^{1/4}\sqrt{3}} \,
\frac{(5-q)^{1/2}\,(3+q)^{1/2}}{2} \,
\left[\frac{(1-q)^{1/2}\,\Gamma\left(\frac{2q}{1-q}+\frac{3}{2}\right)}
{\Gamma\left(\frac{2q}{1-q}+1\right)}
\right]^{1/2}
\end{eqnarray}
if $-1<q<1$.

\begin{figure}[!htb]
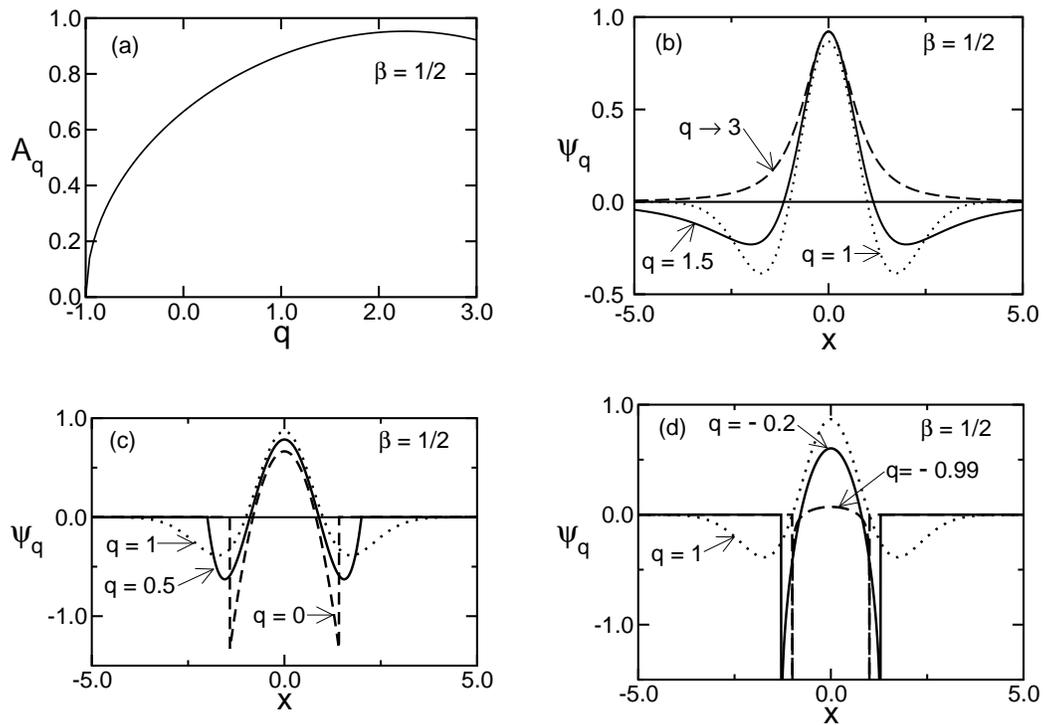

\begin{minipage}[h]{0.47\textwidth}
\epsfig{figure=borgesfig1a.eps,width=\textwidth,clip=}
\end{minipage}
\hfill
\begin{minipage}[h]{0.47\textwidth}
\epsfig{figure=borgesfig1b.eps,width=\textwidth,clip=}
\end{minipage}
\\ $ $  \\ $ $ \\
\begin{minipage}[h]{0.47\textwidth}
\epsfxsize=1.0\textwidth
\epsfig{figure=borgesfig1c.eps,width=\textwidth,clip=}
\end{minipage}
\hfill
\begin{minipage}[h]{0.47\textwidth}
\epsfxsize=1.0\textwidth
\epsfig{figure=borgesfig1d.eps,width=\textwidth,clip=}
\end{minipage}
\caption{$q$-mexican hat with $\beta=1/2$.
(a) Normalization constant $A_q$;
(b) $\psi_q(x)$ for $1<q<3$; (c) $0<q<1$; (d) $-1<q<0$.
The usual mexican hat ($q=1$) is represented by a dotted line,
for comparison.}
\label{Fig:qmex.beta0.5}
\end{figure}

The function $\psi_q(x)$ satisfies the admissibility condition
and consistently recovers the mexican hat, 
$ \lim_{q \rightarrow 1} \psi_q(x) = \psi_1(x).$
The range of admissible values for $q$ $(-1<q<3)$ is divided in
three regions.
For $1<q<3$, $\psi_q(x)$ is infinitely supported
and presents a power-law tail $\sim -1/|x|^{2/(q-1)}$, in marked
contrast with the exponential tail of the original mexican hat.
When $q<1$, a cut-off naturally appears at $|x_c|=[(1-q)\beta]^{-1/2}$.
In the range $0<q<1$, $\psi_q(x_c)=0$, and
when $-1<q<0$, $\psi_q(x_c)$ diverges.
For $q \rightarrow -1$, $\psi_q(x)$ coincides with the abscissa
axis, except at the cut-off positions,
where it diverges. These features introduce significant differences from
the original mexican hat wavelet.
Figure~\ref{Fig:qmex.beta0.5}
illustrates $\psi_q(x)$ with $\beta=1/2$.

The Fourier transform,
\begin{eqnarray}
\label{Fourier}
{\cal F}\left[f(x);\,y\right] \equiv F(y) \equiv
\frac{1}{\sqrt{2\pi}} \int_{-\infty}^{\infty}\rme^{i xy}\,f(x)\,\rmd x,
\end{eqnarray}
of $\psi_q(x)$ may be found by considering (\ref{qsombrero_def}),
and taking into account the property of the Fourier transform of derivatives,
$ {\cal F} \left[f^{(n)};\,y\right]=(-i y)^n\,F(y),$
together with the Fourier transform of a $q$-Gaussian
(see Equations 3.384~9 and 3.387~2 of \cite{Gradshteyn}).
We find for $1<q<3$,
\begin{eqnarray}
{\cal F}\left[\psi_q(x);\,y\right]= \frac{A_q}{(2-q)\beta}
\frac{1}{\sqrt{2(q-1)\beta}\,\Gamma\left(\frac{2-q}{q-1}\right)}
y^2 \left[\frac{|y|}{2\,\sqrt{(q-1)\beta}}\right]^{\nu} 
\nonumber \\
\times
K_{\nu}\left(\frac{|y|}{\sqrt{(q-1)\beta}}\right),
\end{eqnarray}
and for $-1<q<1$,
\begin{eqnarray}
{\cal F}\left[\psi_q(x);\,y\right]= \frac{A_q}{2(2-q)\beta}
\frac{\Gamma\left(\frac{2-q}{1-q}+1\right)}{\sqrt{2(1-q)\beta}}
\; y^2 \left[\frac{2\sqrt{(1-q)\beta}}{y}\right]^{-\nu}
\nonumber \\
\times
J_{-\nu}\left(\frac{y}{\sqrt{(1-q)\beta}}\right),
\end{eqnarray}
with $\nu=\frac{2-q}{q-1}-\frac{1}{2}$.
$J_{-\nu}$ is the Bessel functions
of first kind and $K_{\nu}$ is the modified Bessel function of second kind.

It is possible to have variations of the $q$-mexican hat by using
$\beta=\beta(q)$ (with $\beta(1)=\frac{1}{2}$), for instance, $\beta=1/(3-q)$.

\section{Modulated $q$-Gaussian}
\label{sec:q-gaussian}

\begin{figure}[ht]
\begin{center}
\epsfig{figure=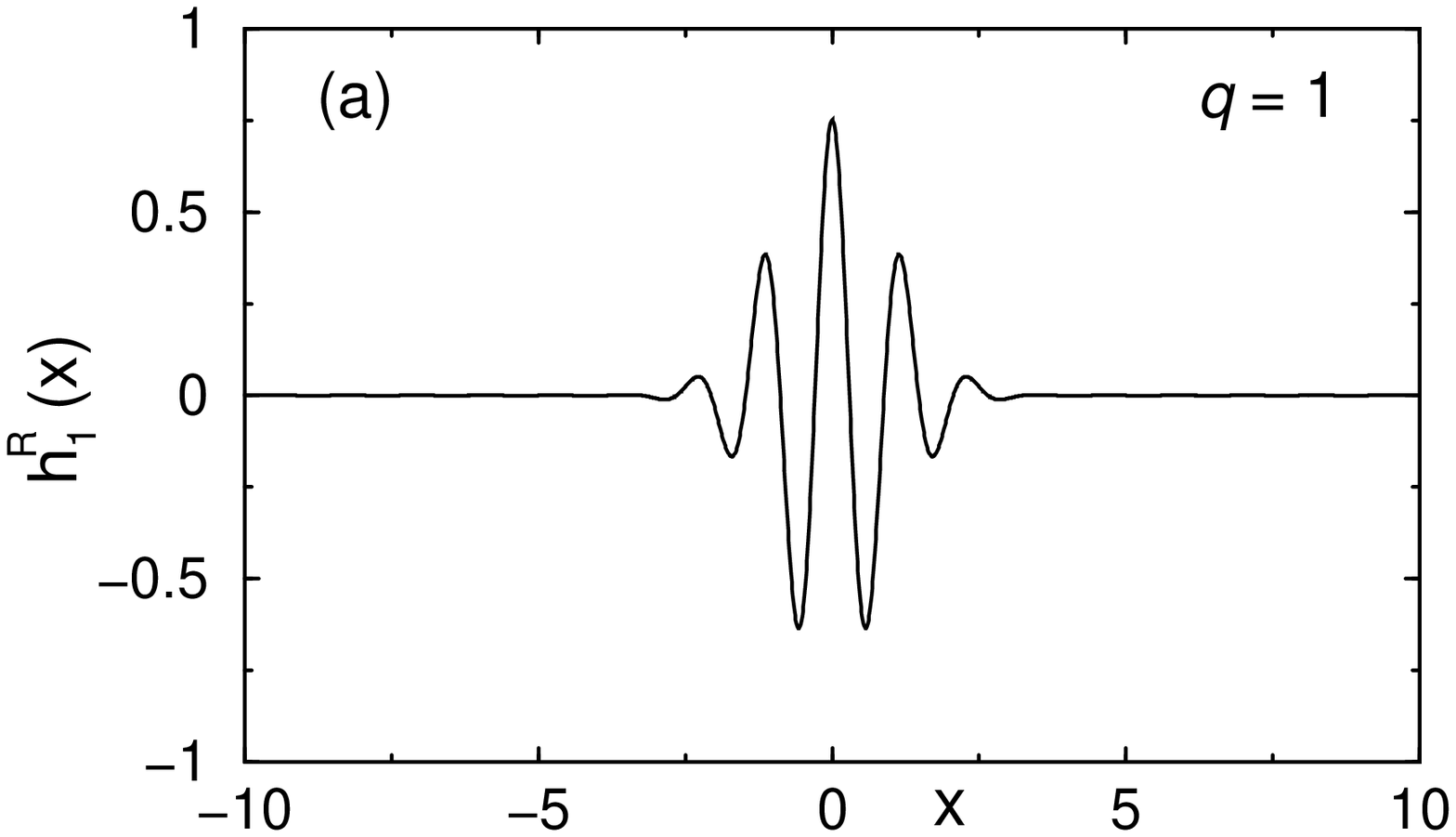,width=0.6\textwidth,clip=}

\epsfig{figure=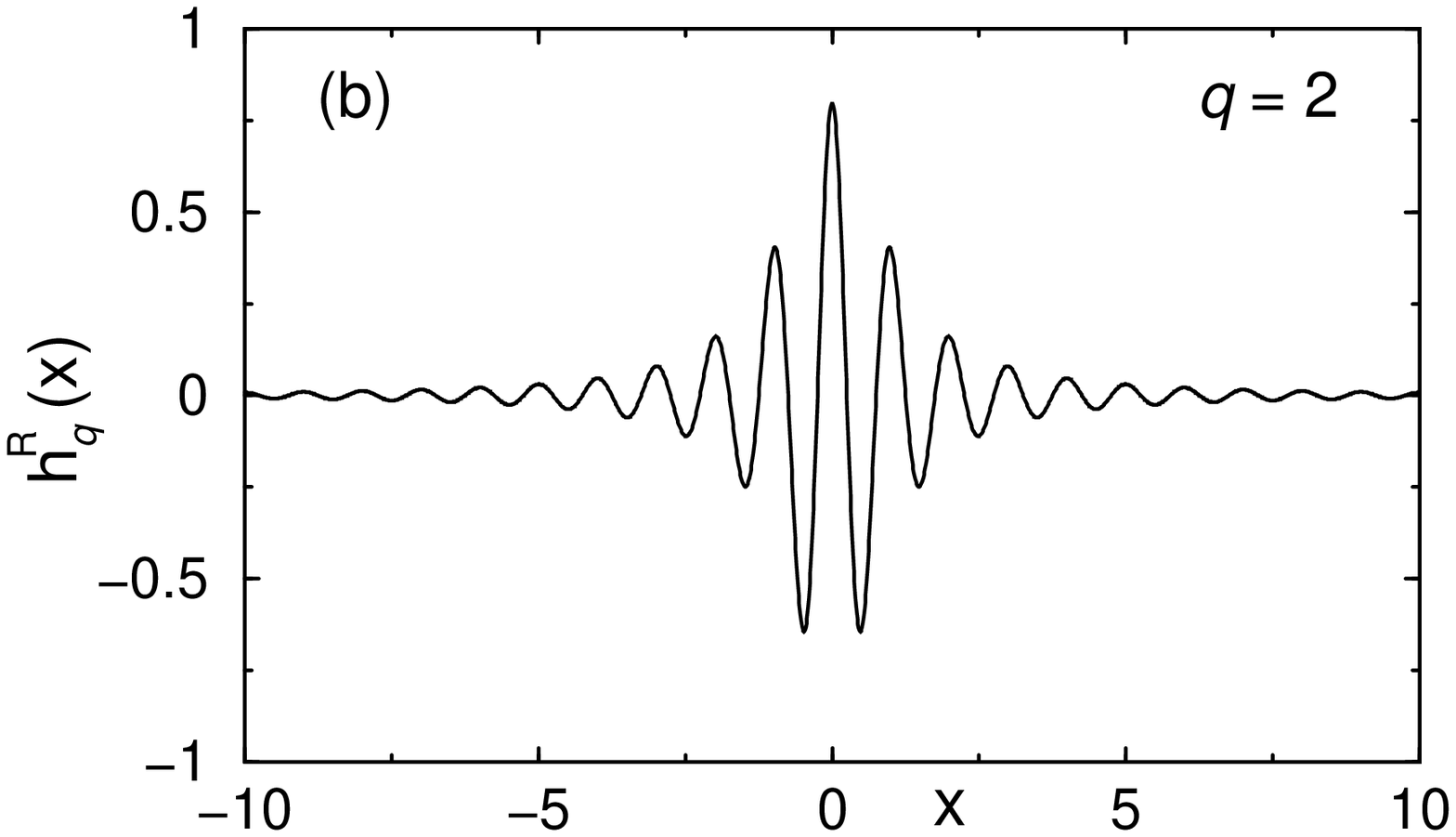,width=0.6\textwidth,clip=}

\epsfig{figure=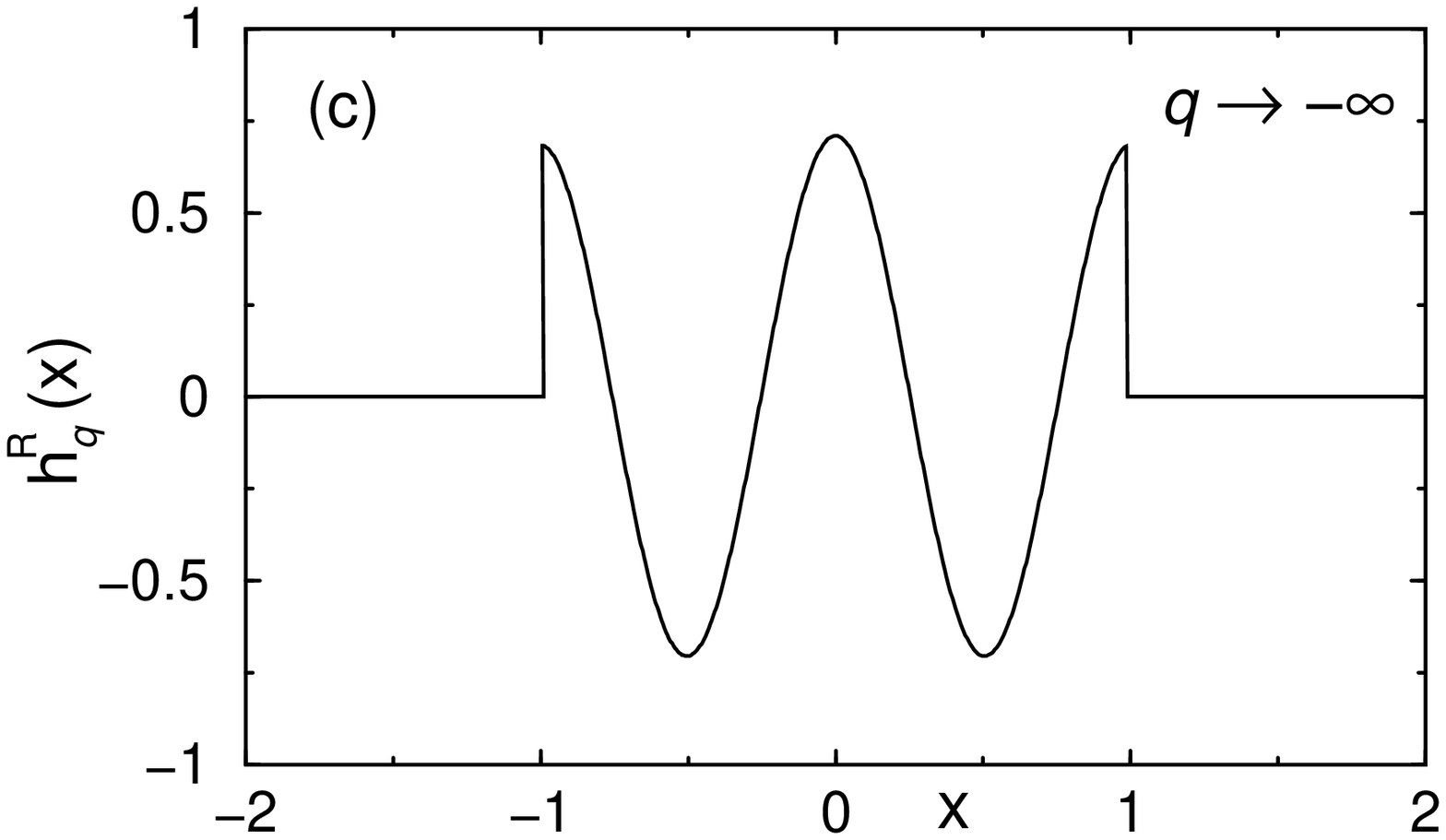,width=0.6\textwidth,clip=}
\end{center}
\caption{Normalized real part of the $q$-modulated Gaussian
($h_q^R(x)\equiv \sqrt{2}\,\mbox{Re }h_q$).
(a) $q=1$ (usual case); (b) $q=2$; (c) $q\to - \infty$
(illustrated with $q=-100$).
Note that abscissa scale in (c) is different from the others.}
\label{Fig:hq}
\end{figure}

Now we turn to the Morlet's wavelet, or modulated Gaussian
\cite{Morlet:82,Grossmann_Morlet:84,Grossmann_Morlet:85},
a function associated with the birth of wavelet analysis
\cite{Temme,Meyer}.
Within the context we are dealing with, we search for a wavelet that is a
generalization of
\cite{Daubechies90,Holschneider}:
\begin{eqnarray}
\label{q=1modgauss}
h(x)= \pi^{-1/4} \, (\rme^{-i kx}-\rme^{-k^2/2}) \, \rme^{-x^2/2},
\end{eqnarray}
where $k=\pi(2/\ln 2)^{1/2}$.
So we simply modulate the usual trigonometric
functions ($\rme^{-i k_qx}$) with a $q$-Gaussian ($\rme_q^{-\beta x^2}$):
\begin{eqnarray}
\label{qmodgauss}
h_q(x)=B_q\,\left(\rme^{-i k_qx}-\Lambda_q(k_q)\right)\,\rme_q^{-\beta x^2},
\quad
\infty < q < 3.
\end{eqnarray}
The function $\Lambda_q(k_q)$ is such that the admissibility condition,
written in the domain of frequencies,
\begin{eqnarray}
{\cal F} [h_q(x);\,0]=0,
\end{eqnarray}
is satisfied. It means that
\begin{eqnarray}
\Lambda_q(k_q) \equiv
\frac{{\cal F} \left[ \rme^{-i k_qx}\,\rme_q^{-\beta x^2};\,0 \right]}
{{\cal F} \left[ \rme_q^{-\beta x^2};\,0 \right]}.
\end{eqnarray}
Taking into account the Fourier transform of a $q$-Gaussian
(see Equations 3.384~9 and 3.387~2 of \cite{Gradshteyn}),
and
${\cal F}[\rme^{-i k_qx};\,y]=\sqrt{2\pi}\,\delta(y-k_q)$,
and also, from the convolution theorem
(with the symmetric convention (\ref{Fourier}) we are adopting here),
\begin{eqnarray}
\label{convolution}
{\cal F} \left[ f(x)\,g(x);\,y \right]
&=& F(y) \ast G(y) \nonumber \\ \noalign{\smallskip}
&=& \frac{1}{\sqrt{2\pi}} \int_{-\infty}^{\infty} F(y-\xi)\,G(\xi) \, \rmd\xi,
\end{eqnarray}
we find for $q>1$,
\begin{eqnarray}
\label{Lambdaq:q>1}
\Lambda_q(k_q)=
\frac{2}{\Gamma\left(\frac{1}{q-1}-\frac{1}{2}\right)}
\left(\frac{k_q}{2\,\sqrt{(q-1)\beta}}\right)^\mu
K_{\mu}\left(\frac{k_q}{\sqrt{(q-1)\beta}}\right),
\end{eqnarray}
and for $q<1$,
\begin{eqnarray}
\label{Lambdaq:q<1}
\Lambda_q(k_q)=
\Gamma\left(\frac{1}{1-q}+\frac{3}{2}\right)
\left(\frac{2\,\sqrt{(1-q)\beta}}{k_q}\right)^{-\mu}
J_{-\mu}\left(\frac{k_q}{\sqrt{(1-q)\beta}}\right),
\end{eqnarray}
with $\mu=\frac{1}{q-1}-\frac{1}{2}$.
We follow the same criterion adopted by \cite{Daubechies90}
for the
determination of the value of $k_q$: the ratio between the second highest
and highest local maxima of $\mbox{Re }h_q$ is $\frac{1}{2}$. It results that
\begin{eqnarray}
\label{k_q}
k_q=2\pi \sqrt{\frac{(q-1)\beta}{2^{q-1}-1}} \qquad
(q \stackrel{\scriptscriptstyle >}{\scriptscriptstyle <} 1).
\end{eqnarray}
The normalization constant $B_q$ is given by
\begin{eqnarray}
\label{B_q}
B_q=
\left\{
\begin{array}{lc}
\left(\frac{\beta}{\pi}\right)^{1/4}
\left[\frac{(q-1)^{1/2}\,\Gamma\left(\frac{2}{q-1}\right)}
{\Gamma\left(\frac{2}{q-1}-\frac{1}{2}\right)}
\right]^{1/2},
& 1<q<3, \\ \noalign{\medskip}
\left(\frac{\beta}{\pi}\right)^{1/4}
\left[\frac{(1-q)^{1/2}\,\Gamma\left(\frac{2}{1-q}+\frac{3}{2}\right)}
{\Gamma\left(\frac{2}{1-q}+1\right)}
\right]^{1/2},
& \qquad q<1.
\end{array}
\right.
\end{eqnarray}
($B_q$ is an approximation, as $B_1=\pi^{-1/4}$ of Equation~(\ref{q=1modgauss})
is also an approximation.)
Figure~\ref{Fig:hq}
illustrates $\sqrt{2}\,\mbox{Re }h_q$ with $\beta(q)=1/(3-q)$.
(The factor $\sqrt{2}$ is used to have the real part normalized.)
We observe the long (power-law) tail for $1<q<3$, in marked contrast with
the rapidly vanishing tail for $q=1$. The cut-off is present for $q<1$.
In the case $q \rightarrow -\infty$, $h_q(x)$ reduces
to a two-cycle function, the imaginary part of it is a variation of
the one-cycle sine presented in \cite{Daubechies90}.

\section{$q$-Trigonometric wavelets}
\label{sec:q-trigonometric}

The $q$-exponential function (\ref{expq}), expanded to the imaginary
(or, more generally, complex) domain by analytic continuation, leads to
$q$-trigonometric functions
\cite{EPB1998}:
\begin{eqnarray}
\label{sinq-cosq}
\cos_q x = \frac{\rme_q^{i x}+\rme_q^{-i x}}{2}, \qquad
\sin_q x = \frac{\rme_q^{i x}-\rme_q^{-i x}}{2i}.
\end{eqnarray}
The $q$-cosine and $q$-sine functions may be expressed as
$ \rme_q^{i x}=\rho_q(x) \, \rme^{i\varphi_q(x)}$
where
\begin{eqnarray}
\rho_q^2(x) = \rme_q^{(1-q)x^2}, \qquad
\varphi_q(x) = \frac{\arctan_1[(1-q)x]}{1-q}.
\end{eqnarray}
We want to construct a wavelet based on such $q$-trigonometric functions.
For this purpose, we recall that the derivative of a $q$-exponential may
be expressed as
\begin{eqnarray}
\frac{\rmd \rme_q^x}{\rmd x} = \rme_{2-1/q}^{qx} \,.
\end{eqnarray}
Once $\rho_{q>1}(x) \rightarrow 0$ for $|x| \rightarrow \infty$,
the admissibility condition
is satisfied for $1<q<2$. Renaming the parameter $q$, we define the following
$q$-trigonometric wavelet:
\begin{eqnarray}
\label{q-trig-wavelet}
\mbox{wt}_q(x) \equiv C_q \, \rme_q^{\frac{i x}{2-q}}, \qquad 1<q<2.
\end{eqnarray}
The normalization constant
is given by:
\begin{eqnarray}
C_q=\sqrt{\frac{1}{2-q}}\,\frac{1}{\pi^{1/4}}
\left[\frac{(q-1)\,\Gamma\left(\frac{1}{q-1}\right)}
{\Gamma\left(\frac{1}{q-1}-\frac{1}{2}\right)}
\right]^{1/2}.
\end{eqnarray}
For brevity of notation, we can write the real and imaginary parts of
$\mbox{wt}_q(x)$ as
\begin{eqnarray}
\mbox{wc}_q (x) \equiv \sqrt{2} \, \mbox{ Re } \mbox{wt}_q(x)
= \sqrt{2} \, C_q\,\cos_q\left(\frac{x}{2-q}\right),
\\ \noalign{\medskip}
\mbox{ws}_q(x) \equiv \sqrt{2} \, \mbox{ Im } \mbox{wt}_q(x)
= \sqrt{2} \, C_q\,\sin_q\left(\frac{x}{2-q}\right).
\end{eqnarray}

\begin{figure}[!htb]
\begin{minipage}[h]{0.48\textwidth}
\begin{center}
\epsfig{figure=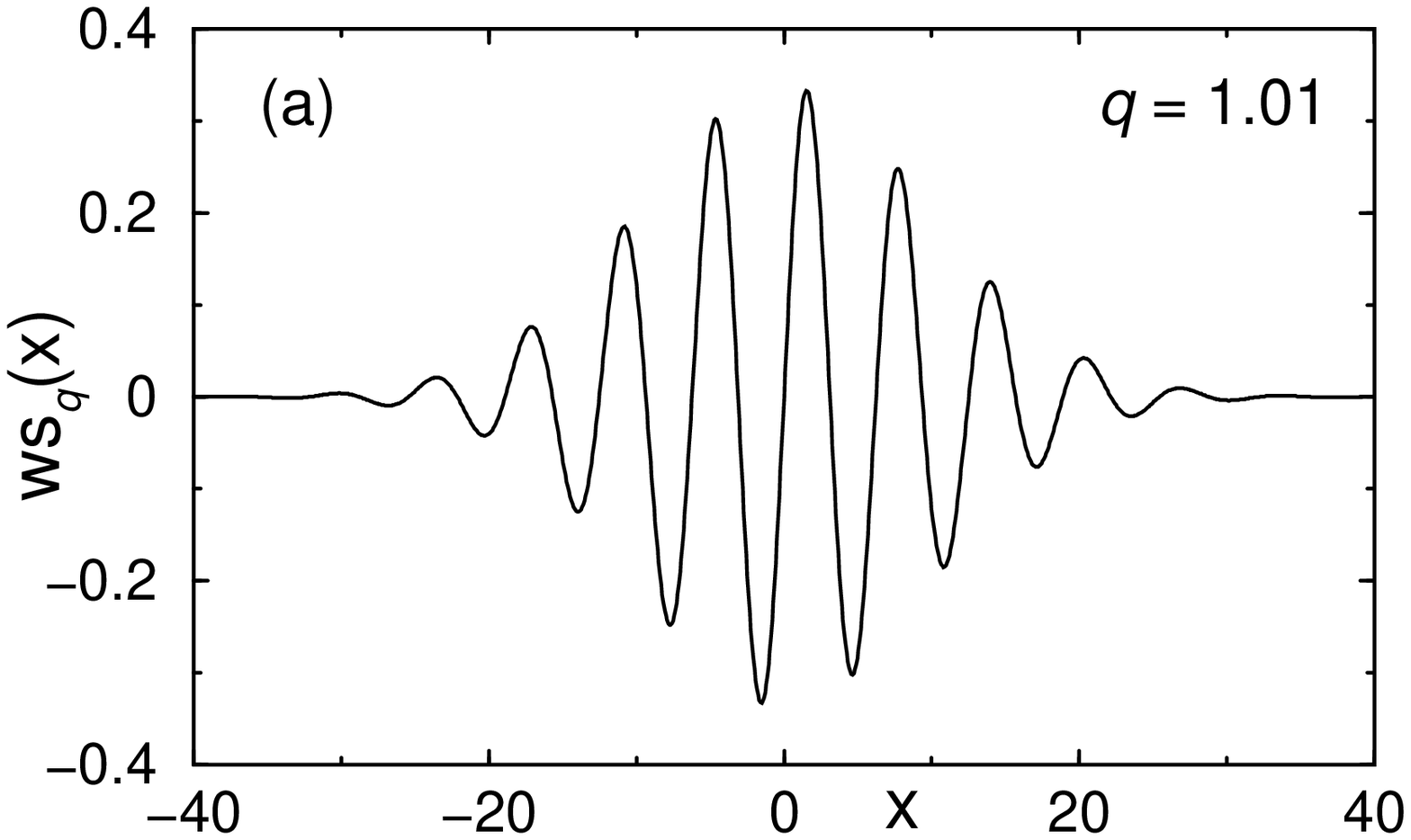,width=\textwidth,clip=}
\end{center}
\end{minipage}
\hfill
\begin{minipage}[h]{0.48\textwidth}
\begin{center}
\epsfig{figure=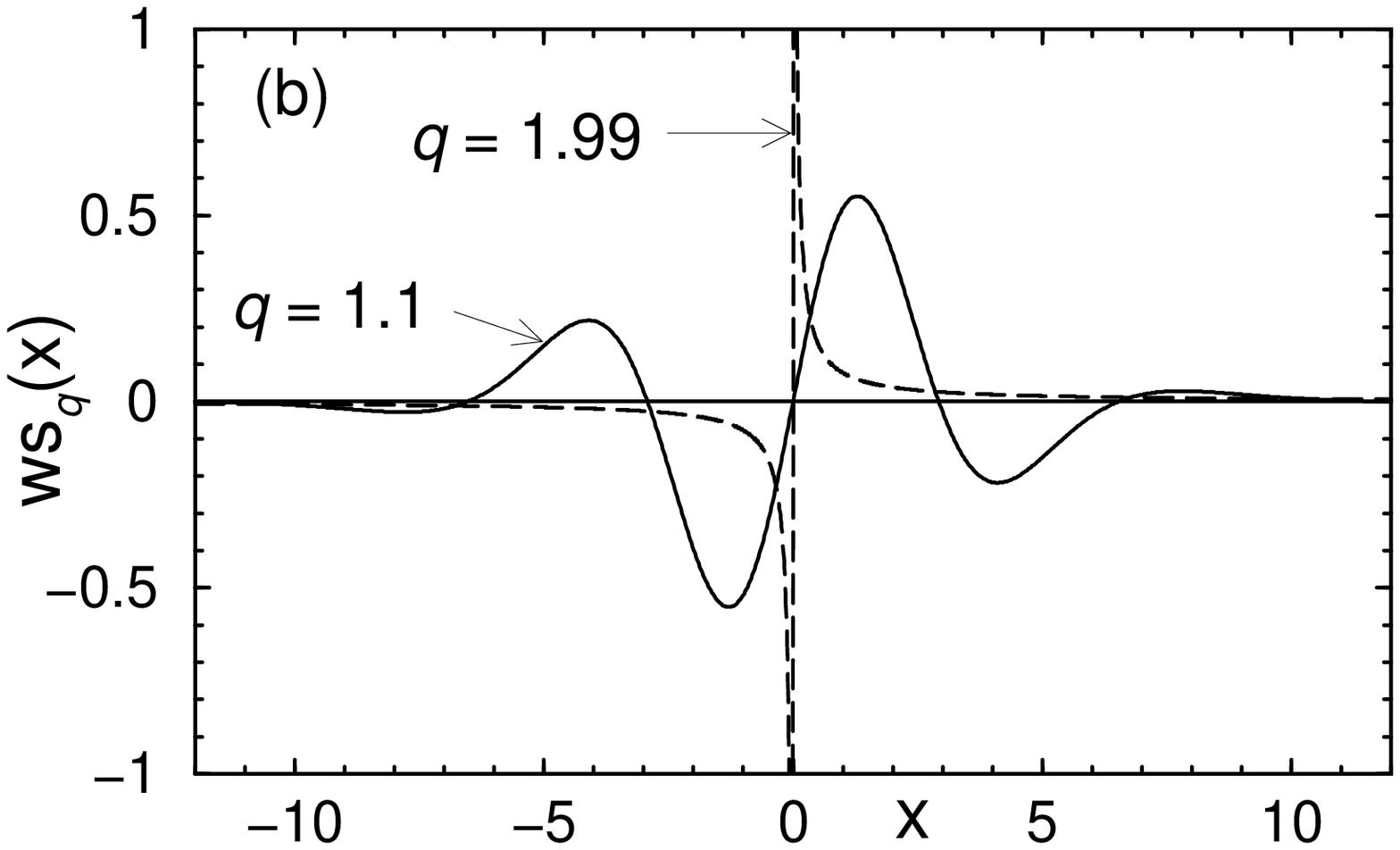,width=\textwidth,clip=}
\end{center}
\end{minipage}
\caption{$\mbox{ws}_q (x)$ vs. $x$. (a) $q=1.01$;
(b) $q=1.1$ (solid) and $q=1.99$ (dashed).}
\label{Fig:wsinq}
\end{figure}

Some typical curves are shown in
Figure~\ref{Fig:wsinq}
(for brevity, we only show the odd function $\mbox{ws}_q(x)$).
Note that the number of oscillations decreases as $q$ goes from $1$ to $2$.
The functions present infinite oscillations of vanishing amplitudes at
$q \rightarrow 1$ ($C_{q \rightarrow 1} \rightarrow 0$).
$\mbox{ws}_q(x)$ presents only one root, at $x_0=0$, 
for $\frac{3}{2}\leq q<2$,
and
$\mbox{wc}_q(x)$ presents only one pair of roots, at
$
x_0=\pm\,\frac{2-q}{q-1}\,\tan\left[(q-1)\frac{\pi}{2}\right],
$
for $\frac{4}{3}\leq q<2$.
As $q \rightarrow 2$, $C_q \rightarrow \infty$, and
the wavelets become sharply localized
(the roots of $\mbox{wc}_q(x)$ approaches $\pm\,2/\pi$).
Let us also mention that the modulation of the functions is not Gaussian,
but rather a power-law decay. But they are essentially different from
the modulated $q$-Gaussian derived in the previous Section.

The Fourier transform of $\mbox{wc}_q (x)$ and $\mbox{ws}_q (x)$ are found
with Equation~(\ref{sinq-cosq}) and
(see Equations 3.382~6 and 3.382~7 of \cite{Gradshteyn})
\begin{eqnarray}
\label{Fourier e_q(iax)}
{\cal F}\left[\rme_{q>1}^{i ax};\,y\right]=
\left\{
\begin{array}{ll}
0\;, & y>0, \\ \noalign{\medskip}
\displaystyle
\frac{\sqrt{2\pi}\,(-y)^{\frac{1}{q-1}-1}\,\rme^{\frac{y}{(q-1)a}}}
{[(q-1)a]^{\frac{1}{q-1}}\,\Gamma\left(\frac{1}{q-1}\right)}
,
& y<0,
\end{array}
\right.
\end{eqnarray}
and
\begin{eqnarray}
\label{Fourier e_q(-iax)}
{\cal F}\left[\rme_{q>1}^{-i ax};\,y\right]=
\left\{
\begin{array}{ll}
\displaystyle
\frac{\sqrt{2\pi}\,y^{\frac{1}{q-1}-1}\,\rme^{\frac{-y}{(q-1)a}}}
{[(q-1)a]^{\frac{1}{q-1}}\,\Gamma\left(\frac{1}{q-1}\right)}
,
& y>0, \\ \noalign{\medskip}
0\;, & y<0.
\end{array}
\right.
\end{eqnarray}
%

\section{An Example}
\label{sec:example}

Wavelet transforms constitute a multi-purpose tool that rapidly met
wide\-spread applications in many areas of pure and applied sciences. To
exemplary demonstrate the reliability of $q$-wavelets, we briefly present how
they can be used to reproduce several properties of well known fractal sets.

\begin{figure}[!htb]
\begin{center}
\epsfig{figure=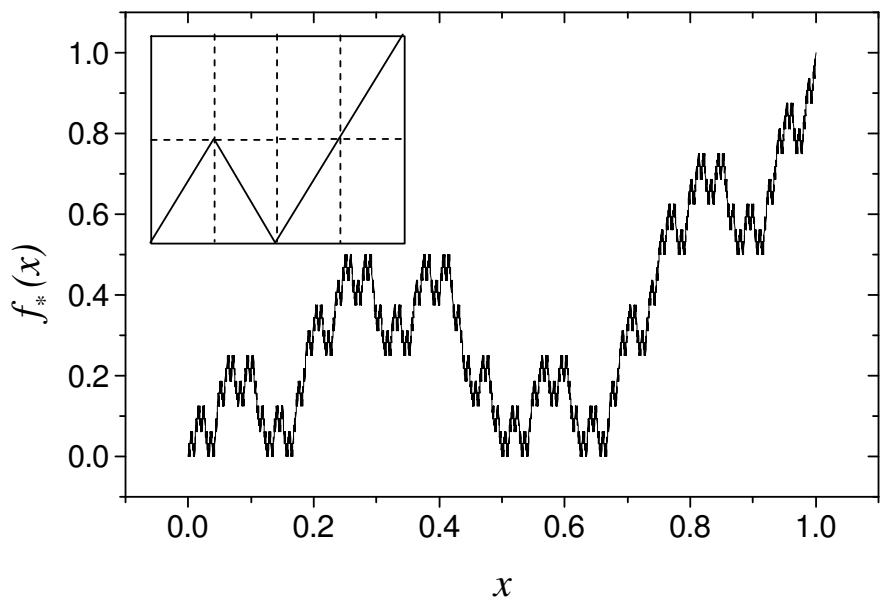,width=\textwidth,clip=}
\end{center}
\caption{Self-affine monofractal profile. Inset: its generator.}
\label{Fig:profile}
\end{figure}

Consider a self-affine profile $y=f_{*}(x)$ shown in Figure~\ref{Fig:profile}.
Its construction starts with the generator shown in the Inset.
The generator is recursively applied for each straight line segment.
The self-affine fractal profile is obtained in the limit of
infinite iterations.

The resulting figure after infinite generations has a local fractal
dimension $D=2-\alpha ,$ where $\alpha =\ln B_y/\ln B_x=0.5$ is the
roughness exponent, and $B_x$ and $B_y$ are the scaling factors along
the $x$ and $y$ axis respectively.
To analyze the scaling properties of $f(x)$ around an
arbitrary point $x_0,$ we shift the origin and define

\begin{equation}
\label{w0}
f_{x_0}(x)=f(x_0+x)-f(x_0)\approx Ax^{\alpha (x_0)}
\end{equation}
where $\alpha (x_0),$ the Holder (or singularity) exponent of $f_{x_0}(x),$
indicates how this function vanishes (or diverges) at $x=x_0.$

Using a general wavelet transform of $f_{x_0}(x),$ it is straightforward to
show that the Holder exponent follows immediately from the wavelet transform
as
\begin{equation}
\label{w1}
T_g(a,b)\;f_{x_0}(x)\sim a^{\alpha (x_0)}.
\end{equation}

\begin{figure}[!htb]
\begin{center}
\epsfig{figure=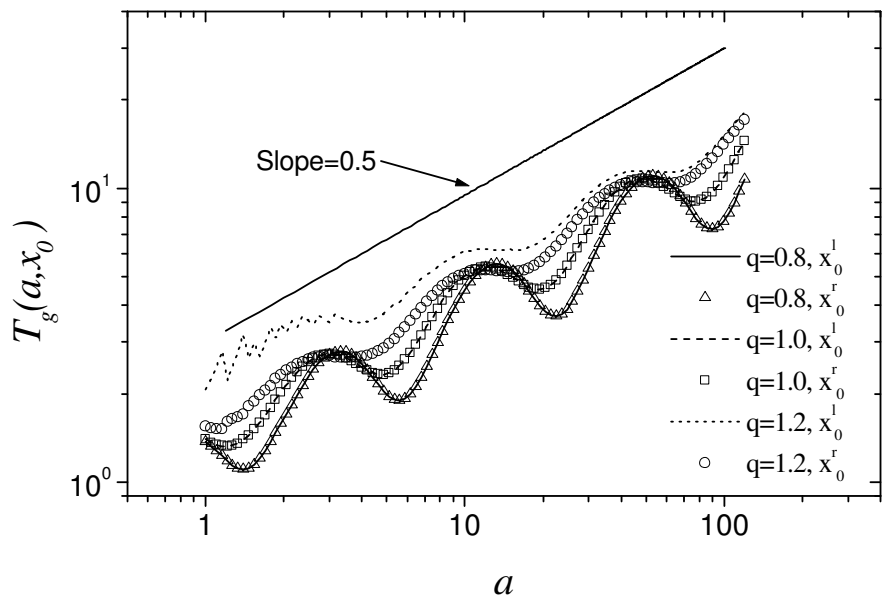,width=\textwidth,clip=}
\end{center}
\caption{Wavelet transform (Equation (\protect\ref{w1}))
with two different values of $b$: $x^l_0=0.25$ and $x^r_0=0.50$.
Note that the curves for the same values of $q$ coincide, as it
should be for a monofractal.}
\label{Fig:monofractal}
\end{figure}

Figure~\ref{Fig:monofractal} shows how the $q$-wavelet transform
(Equation (\ref{w1})) behaves for two points along the profile for some
values of $q.$ The curves clearly indicates that the value $\alpha=0.5$
is accurately reproduced. We observe that the results for
$q>1$ show small irregularities. This is due to the fact that, as
$\rme_q^{-x}$ decays only as a power-law when $x\rightarrow \infty$,
the numerical integration in (\ref{w1}) must be performed over a
wider interval than for $q\leq 1.$ In order to make evident this
effect, we have performed all integrals over a fixed integration interval.
On the other hand, results for $q<1$ show good convergence, as the function
naturally presents a cutoff. Oscillations in the curves are typical for
this kind of analysis, as shown in the same figure for the usual mexican hat
($q=1$).

\begin{figure}[!htb]
\begin{center}
\epsfig{figure=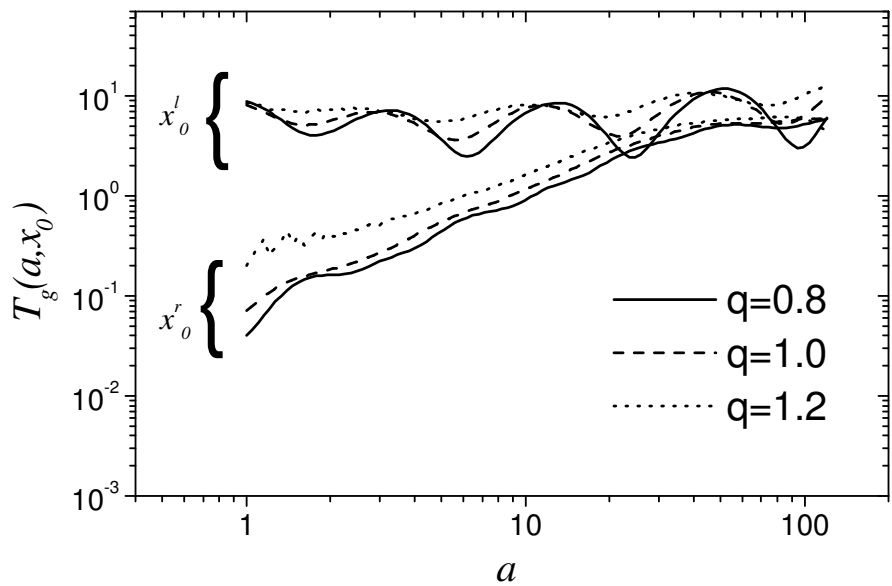,width=\textwidth,clip=}
\end{center}
\caption{Wavelet transform (Equation (\protect\ref{w1}))
with two different values of $b$: $x^l_0=0.25$ (upper curves)
and $x^r_0=0.75$ (lower curves).
The slopes are clearly different, as a signature of the
multifractal nature.}
\label{Fig:multifractal}
\end{figure}

Results for a more complex situation can be explored if we
consider a multifractal set. It is generated along a similar way
used to obtain the first set, where we choose two different
scaling factors for the first and second half of the profile in
each generation of its construction. The plots for different
values of $x_0$ have different slopes (Figure~\ref{Fig:multifractal}),
indicating that many scaling laws and fractal dimensions are found in the
resulting profile.

$q$-Wavelets also prove to be a reliable tool to analyze this much more
complex situation that requires a multi-fractal formalism. This amounts to
evaluate the generalized fractal dimensions $D_Q,$ or its Legendre
transformed singularity spectrum $f(\alpha )=\tau (q)-\alpha Q$, where
$\alpha $ represents the Holder exponent. This analysis proceeds
within the so-called Wavelet Transform Modulus-Maxima Method
(WTMMM), that has been developed in recent years. We will not show
the details that can be found, e.g., in \cite{Arneodo}.

\begin{figure}[!htb]
\begin{center}
\epsfig{figure=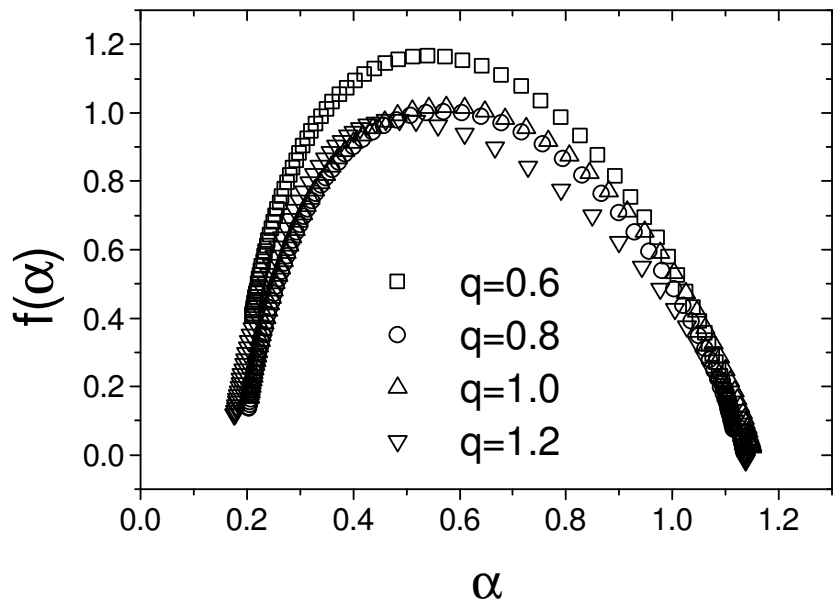,width=\textwidth,clip=}
\end{center}
\caption{$f(\alpha)$ spectra for the multifractal profile.
For $q=0.8$ results are very similar to those with $q=1$. $q=0.6$
yields spurious results.}
\label{Fig:f(a)}
\end{figure}

Figure~\ref{Fig:f(a)} shows the $f(\alpha)$ spectra, for
different values of $q$, corresponding to the multifractal
profile. The graphs indicate that the spectra produced by
$q$-wavelets, with $q<1$, are comparable to the one obtained by
the usual $q=1$ mexican hat. As we fixed the integration interval
for all values of $q,$ we observe major differences in the spectra
for values $q>1.$ This limitation to its practical use is
expressed by larger numerical effort to compute the integrals in
the wavelet transforms with the same accuracy for values of $q\leq 1$.

\section{Final Remarks}
\label{sec:final}

The functions here introduced generalize widely used mother wavelet functions
by means of a single parameter.
The $q$-wavelets present significant different behaviours, when compared
with the original $q=1$ cases. The fingerprints of the generalized
$q$-wavelets are the power-law decay $(q>1)$, and the cut-off $(q<1)$.
The generalization was inspired in the nonextensive statistical mechanics
and the $q$-exponential that emerges from it. Within this formalism,
the entropic index $q$ measures departures from the usual
Boltzmann-Gibbs behaviour.
There is a number of possible origins for such departures,
like slow (power-law) mixing, long-range (spatial) interactions,
long-term (temporal) memory or fractal nature of the phase space.
When dealing with signals which exhibit some of these features,
one can hopefully take advantage of the index $q$ to tune the
wavelet to the signal, as it is here briefly exemplified for the
$q$-mexican hat. Other wavelets, as well as scaling functions, may
be generalized along the lines here introduced, for instance, the
Meyer wavelet or the sinc function.

\section{Acknowledgments}

We thank C Rodrigues Neto for his collaboration at the early stages of this
project, and E~K Lenzi for useful remarks.
This work is partially supported by PRONEX/MCT, CNPq, CAPES and FAPERJ
(Brazilian agencies).


%
%

%
%

%
%

%
%

%
%

%
%

%
%

\end{document}